\documentclass[runningheads]{llncs}
\usepackage[T1]{fontenc}
\usepackage{physics}
\usepackage{bbm}
\usepackage{bbold}
\usepackage{amsmath}
\usepackage{amssymb}
\usepackage{amsfonts}
\usepackage{graphicx}
\usepackage{hyperref}
\usepackage[dvipsnames]{xcolor}
\usepackage{todonotes}
\usepackage{booktabs}

\newcommand{\vect}[1]{\boldsymbol{#1}}

\newcommand{\1}{\mathbb{1}}
\newcommand{\C}{\mathbb{C}}
\begin{document}
\title{Quantum-aware Transformer model for state classification}
% OLD: Quantum-LLM(Transformer): quantum-aware model for state classification
\titlerunning{QLLM}
\author{Przemys{\l}aw Seku{\l}a\inst{1,2}\orcidID{0000-0002-4599-1077} \and
Micha{\l} Romaszewski\inst{1}\orcidID{0000-0002-8227-929X} \and
Przemys{\l}aw G{\l}omb\inst{1}\orcidID{0000-0002-0215-4674} \and
Micha{\l} Cholewa\inst{1}\orcidID{0000-0001-6549-1590}\and
{\L}ukasz Pawela\inst{1}\orcidID{0000-0002-0476-7132}}
\institute{Institute of Theoretical and Applied Informatics, Polish Academy of
Sciences, Ba{\l}tycka 5, 44-100 Gliwice, Poland\\
\and
University of Maryland, College Park, Department of Civil and Environmental 
Engineering, MD, USA \\
\email{\{psekula,mromaszewski,przemg,mcholewa,lpawela\}@iitis.pl}
}

\maketitle              % typeset the header of the contribution
\begin{abstract}
%The abstract should briefly summarize the contents of the paper in 150--250 words.
Entanglement is a fundamental feature of quantum mechanics, playing a crucial role in quantum information processing. However, classifying entangled states, particularly in the mixed-state regime, remains a challenging problem, especially as system dimensions increase. In this work, we focus on bipartite quantum states and present a data-driven approach to entanglement classification using transformer-based neural networks. Our dataset consists of a diverse set of bipartite states, including pure separable states, Werner entangled states, general entangled states, and maximally entangled states. We pretrain the transformer in an unsupervised fashion by masking elements of vectorized Hermitian matrix representations of quantum states, allowing the model to learn structural properties of quantum density matrices. This approach enables the model to generalize entanglement characteristics across different classes of states. Once trained, our method achieves near-perfect classification accuracy, effectively distinguishing between separable and entangled states. Compared to previous Machine Learning, our method successfully adapts transformers for quantum state analysis, demonstrating their ability to systematically identify entanglement in bipartite systems. These results highlight the potential of modern machine learning techniques in automating entanglement detection and classification, bridging the gap between quantum information theory and artificial intelligence.

\end{abstract}
\section{Introduction}
The entanglement phenomenon is at the heart of quantum information, enabling its key applications such as quantum computing, secure communication, and enhanced metrology. Unlike classical correlations, entanglement represents a uniquely quantum feature where the state of a system cannot be described independently of its subsystems. This fundamental property underlies protocols like quantum teleportation, superdense coding, and quantum key distribution, as well as quantum computational speedups~\cite{Watrous2018theory,Nielsen2010quantum}. However, not all quantum states exhibit entanglement, and distinguishing entangled states from separable ones is a crucial yet challenging problem in quantum information science. The ability to efficiently classify quantum states has direct implications for the practical implementation of quantum technologies, motivating the development of reliable entanglement detection and classification methods.

Bipartite quantum states, which describe systems naturally partitioned into two subsystems, are fundamental in quantum information science and serve as a basis for studying quantum correlations and computational advantages. These states reside in a tensor product space  $\mathrm{L}(\C^{d_1} \otimes \C^{d_2})$, where entanglement emerges as a crucial resource for various quantum information applications~\cite{Horodecki2009,Gisin2007}, such as secure communication, quantum-enhanced computation, and efficient information transfer. For pure bipartite states, i.e., vectors in $\C^{d_1} \otimes \C^{d_2}$, entanglement can be clearly identified: a state is separable if and only if it can be expressed as a tensor product of subsystem states. Any departure from this structure signifies entanglement, which directly influences quantum nonlocality, measurement-based quantum computing, and the efficiency of entanglement-assisted protocols.

\begin{figure}
\centering\includegraphics{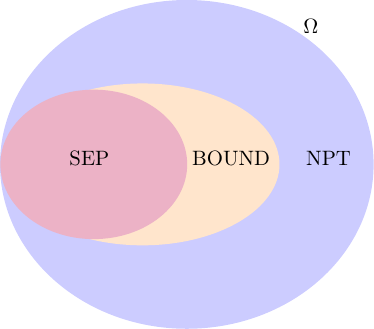}
\caption{Schematic representation of the set of mixed quantum states $\Omega$
depicting separable states (SEP) $\rho_\mathrm{sep}$, bound entangled states
(BOUND) $\sigma_\mathrm{bound}$ and negative partial transpose states (NPT)
$\xi_\mathrm{npt}$.}
\label{fig:set-of-states}
\end{figure}

The identification of entanglement for bipartite states is straightforward in the case of pure states but becomes significantly more challenging when considering mixed bipartite states. 
A variety of analytical and numerical methods have been
developed for the characterization and detection of mixed-state entanglement.
One of the most celebrated approaches is based on the \emph{positivity of the
	partial transpose (PPT)}, introduced by Peres~\cite{Peres1996} and
Horodecki~\cite{Horodecki1996}. While the PPT criterion is both necessary and
sufficient for separability in low-dimensional cases ($\C^2 \otimes \C^2$ and $\C^2 \otimes \C^3$), in higher dimensions, a state can be \emph{PPT and still entangled}. Such
states are known as \emph{bound entangled} states~\cite{Horodecki1998} (see Fig.~\ref{fig:set-of-states} for sketch). They
cannot be distilled into pure entangled states using local operations and
classical communication (LOCC), rendering them entangled yet practically
``inaccessible'' for certain quantum information protocols~\cite{Bennett1999}.
This discovery highlighted the limitations of the PPT criterion and the
intricate nature of entanglement in higher-dimensional systems. To emphasize
this difference, entangled states which are not bound entangled are sometimes
called free entangled states.

To address these limitations, other techniques---particularly
\emph{entanglement witnesses}---have been
introduced~\cite{Horodecki2009,Guhne2009}. An entanglement witness is a
Hermitian operator $W$ with the property that
$\Tr(W\,\rho_{\mathrm{sep}}) \ge 0$ for all separable states
$\rho_{\mathrm{sep}}$, but $\mathrm{Tr}(W\,\rho_{\mathrm{ent}}) < 0$ for at
least one entangled state $\rho_{\mathrm{ent}}$. Finding and optimizing
entanglement witnesses can often be formulated via semidefinite programming
techniques, and in many cases, witnesses can be tailored to detect specific
classes of entangled states, including those exhibiting bound entanglement.
Additional approaches to mixed-state entanglement include various
\emph{entanglement measures} (e.g., negativity, entanglement of formation),
which attempts to quantify the degree of entanglement in a given density
operator~\cite{Vidal2002,Plenio2007}. 

Another class of approaches relies on machine learning (ML) to identify 
entanglement directly from the data. In~\cite{Goes2021Automated}, authors 
use automated ML for state classification. Instead of directly measuring 
entanglement properties, the state is reconstructed, and entanglement is inferred 
from the data itself. 
Recently, Transformers have also been applied to quantum random number 
validation~\cite{Goel2024}, showcasing their ability to handle large input 
sequences efficiently and perform multiple statistical tests in parallel. 
The same self-attention mechanism that captures subtle global dependencies in 
random bit streams can likewise model the intricate correlations of bipartite 
quantum states, suggesting that Transformers are a promising architecture 
for entanglement classification under partial or noisy data.

Despite significant progress, a complete classification of bipartite entangled 
mixed states remains an open challenge, particularly as system dimensions grow. 
In this work, we take a data-driven approach to this problem by generating a 
diverse dataset of pure and mixed states through multiple methods and employing 
transformer-based neural networks to analyze their properties and classify 
them. We demonstrate that entanglement identification can be performed 
effectively from the data itself, extending the range of successful 
classification beyond previous studies. 
Furthermore, we validate the application of transformer architectures in this 
domain, achieving a breakthrough where prior deep learning 
approaches~\cite{Goes2021Automated} have struggled. By integrating machine 
learning with established theoretical criteria, we provide a scalable framework 
for systematically detecting and categorizing entanglement, bridging the gap 
between quantum information theory and modern Artificial Intelligence 
techniques.

This paper is organized as follows. In Section~\ref{sec:dataset}, we explain the methodology for constructing our dataset. In Section~\ref{sec:transformer}, we detail our proposition of the Quantum-aware Transformer model and its training scheme. In Section~\ref{sec:rd}, we present the results of validation experiments and discussion.

The code for this paper is publicly available on GitHub\footnote{\url{https://github.com/iitis/LQM}} 
under an open license to facilitate result reproducibility and transparency
and the data can be shared upon a reasonable request.

\section{Dataset generation}\label{sec:dataset}

Quantum states, the basic objects of quantum mechanics, can be broadly
classified as \emph{pure} or \emph{mixed}. A \emph{pure state} is described by a
single $d$-dimensional vector $\ket{\psi}$ in a complex Euclidean space $\C^d$,
and a corresponding density operator $\rho = \ketbra{\psi}{\psi}$. This operator
satisfies $\rho^2 = \rho$ and $\mathrm{Tr}(\rho) = 1$. When $d=2$, the corresponding
system is called a qubit; for $d=3$, it is called a qutrit.

In contrast, a \emph{mixed state} is represented by a statistical ensemble of
pure states, described by a density operator $L(\C^d)\ni\rho = \sum_i p_i
\ketbra{\psi_i}{\psi_i}$, where $p_i \in [0, 1]$ and $\sum_i p_i =
1$~\cite{Nielsen2010}. Such states often emerge from partial traces of larger
systems or incomplete information about the quantum system under study.
Note that $\rho$ is a positive semidefinite matrix.

In many quantum information scenarios, one focuses on \emph{bipartite} states.
These describe a physical system that can be naturally partitioned into two
subsystems, $\ket{\psi} \otimes \ket{\phi}$, associated with complex Euclidean
spaces $\C^{d_1}$ and $\C^{d_2}$. The total state then lives in the tensor
product space $\C^{d_1} \otimes \C^{d_2}$. Bipartite quantum systems are not
only a cornerstone for foundational studies of quantum correlations but also
play a central role in key quantum information tasks such as quantum
teleportation, quantum key distribution, and superdense coding.

For \emph{pure bipartite states},
there exists a straightforward way to distinguish entangled from separable
states: a pure bipartite state $\ket{\psi} \in \C^{d_1} \otimes \C^{d_2}$ is
separable if and only if it can be written as $\ket{\xi} = \ket{\psi} \otimes
\ket{\phi}$. Any deviation from this product structure indicates entanglement.

In this work, we study the following cases of discrimination between separable and entangled states:
\begin{enumerate}
	\item two-qubit states, $\C^2\otimes \C^2$,
	\item qubit-qutrit systems, $\C^2 \otimes \C^3$,
	\item qutrit-qutrit systems, $\C^3 \otimes \C^3$,
\end{enumerate}

For each of these, we generate a dataset consisting of:
\begin{enumerate}
	\item pure separable states,
	\item general entangled states,
	\item Werner entangled states (for qubit-qubit and qutrit-qutrit systems),
	\item maximally entangled states (for qubit-qubit and qutrit-qutrit systems),
	\item bound entangled states from the family by
	Horodecki~\cite{horodecki1999bound} (for qutrit-qutrit systems).
\end{enumerate}
For this  work we assume we have access to the full
tomography~\cite{cramer2010efficient} of each $\rho$, hence we encode each state
as a vector of $2d_1d_2$ real variables.

The details of the sampling are described in the following subsections. Uniform
sampling of quantum states (either pure or mixed) was conducted utilizing the
QuantumInformation.jl package~\cite{Gawron2018}.

\subsection{Sampling pure separable states}

In this case, we need to sample uniformly normalized vectors of the form $
\C^{d_1} \otimes \C^{d_2} \ni \ket{\psi} = \ket{\phi_1} \otimes \ket{\phi_2}$.
This is done in the following steps:
\begin{enumerate}
\item Sample a non-normalized $\ket{x} \in \C^{d_1}$ with each element $x_i$
such that $\Re(x_i) \thicksim N(0, 1)$ and $\Im(x_i) \thicksim N(0, 1)$.
\item Normalize $\ket{x}$: $\ket{\phi_1} = \frac{\ket{x}}{\| \ket{x} \|}$.
\item Repeat steps 1 and 2 to obtain $\ket{\phi_2}$.
\item Put $\ket{\psi} = \ket{\phi_1} \otimes \ket{\phi_2}$.
\end{enumerate}
This procedure ensures that each $\ket{\phi_i}$ is sampled from the Haar measure.

\subsection{Sampling Werner entangled states}
A Werner state is a mixed state having the form
\begin{equation}
\rho_\mathrm{wer} = (1-p) \ketbra{\psi}{\psi} + p \rho^*,
\end{equation}
where $rho^*$ is the maximally mixed state 
\begin{equation}
	\rho^* = \frac{\1}{d^2}
\end{equation}
and
\begin{equation}
	\ket{\psi} = \frac{1}{\sqrt{d}} \sum_{i=0}^{d-1} \ket{ii}.
\end{equation}	
It can be shown that the state
remain entangled for 
\begin{equation}
	p < \frac{d}{d+1}.
\end{equation}
We sample $p$ uniformly in this interval and construct $\rho_\mathrm{wer}$.

\subsection{Sampling general entangled states}\label{sec:general-ent}
We sample general entangled states by uniformly sampling the state of all mixed quantum states 
and only accepting the state as entangled when it is NPT. The steps are:
\begin{enumerate}
	\item Sample a square Ginibre matrix, $G$, of dimension $d_1 d_2$. Elements
	$G_{ij}$ pertain to the complex normal distribution $\Re(G_{ij}) \thicksim N(0,
	1)$ and $\Im(G_{ij}) \thicksim N(0, 1)$.
	\item Calculate $W=GG^\dagger$.
	\item Normalize the trace, $\rho = \frac{W}{\Tr W}$.
	\item Check the Peres-Horodecki criterion. If $\rho$ is NPT, accept it into the
	set; otherwise, repeat the procedure.
\end{enumerate}
Note that this procedure is quite efficient, especially as the dimension
increases, as the relative volume of the separable states
diminishes~\cite{zyczkowski1999volume,zyczkowski1998volume2,zyczkowski1998volume}.

\subsection{Sampling maximally entangled states}
We sample maximally entangled states by sampling unitary matrices, vectorizing
them, and renormalizing them~\cite{PhysRevA.92.012304}. The procedure is:
\begin{enumerate}
	\item Sample a square Ginibre matrix $G$ of dimension $d$ as described in
	Section~\ref{sec:general-ent}.
	\item Calculate its QR decomposition $G = QR$, where $Q$ is a unitary matrix and
	$R$ is upper triangular.
	\item Multiply $i$-th column of $Q$, $Q_i$, by the phase of the corresponding
	diagonal element of $R$, $R_{ii}$, thus obtaining the $i$-th column of a unitary
	matrix $U$. This step is necessary to ensure the proper distribution of
	eigenvalues of $U$~\cite{kukulski2021generating,ozols2009generate}.
	\item Vectorize the matrix $U$ and normalize by $\frac{1}{\sqrt{d}}$.
\end{enumerate}

\subsection{Sampling bound entangled states}
This procedure is based on a family of bound entangled states described
in~\cite{horodecki1999bound}. First, let us introduce
\begin{equation}
\begin{split}
\sigma_+ &= \frac{1}{3} \left( \ketbra{01} + \ketbra{12} + \ketbra{20} \right) \\
\sigma_- &= \frac{1}{3} \left( \ketbra{10} + \ketbra{21} + \ketbra{02} \right) \\
\ket{\psi} &= \frac{1}{\sqrt{3}} \left( \ket{00} + \ket{11} + \ket{22} \right).
\end{split}
\end{equation}
We construct a state $\rho$ as follows
\begin{equation}
\rho_\alpha = \frac{2}{7} \ketbra{\psi} + \frac{\alpha}{7} \sigma_+ + \frac{5 - \alpha}{7} \sigma_-.
\end{equation}
Depending on parameter $\alpha$ this state can be separable ($2 \leq \alpha \leq
3$), bound entangled ($3 < \alpha \leq 4$) or free entangled ($4 < \alpha \leq
5$). Hence, we sample $\alpha$ uniformly in the range $(3, 4]$.

\section{Quantum-aware Transformer model}\label{sec:transformer}

We use a Transformer~\cite{attention_is_all_you_need} based model for processing 
quantum state matrices, leveraging its self-attention mechanism for structured 
data reconstruction. Transformers, originally designed for language sequence 
modeling, operate on sets of input tokens with global context awareness. 
Here, we treat quantum state matrices as tokenized inputs, where each matrix 
element (real and imaginary parts) forms a structured sequence. 
By applying a masked autoencoding strategy, we train the model to reconstruct 
missing matrix elements, allowing it to learn intrinsic quantum state 
properties. This approach enables the Transformer to capture patterns in 
quantum data and improve downstream tasks like entanglement classification.

\subsection{Model Description}
We employ a masked Transformer architecture, referred to as the 
\emph{MaskedTransformer}, to reconstruct partially masked 2D quantum data. 
The input is first flattened from $\left[B, 2N^2\right]$ into 
$\left[B, N^2, 2\right]$, where $B$ stands for the size of data batch,
$N$ is the size of the original square matrix (each entry contains real and 
imaginary parts). 
Tokens in this representation are randomly masked and replaced by a 
\emph{mask token}, and the partially masked sequence is passed through a 
Transformer encoder composed of multi-head self-attention layers and 
feed-forward networks.

Each token embedding is further augmented with a trainable positional vector 
$\mathbf{p}_i$, ensuring that the Transformer retains the relative location 
of each matrix element in the spatial grid. Concretely, we added $\mathbf{p}_i 
\in \mathbb{R}^{d}$ to the embedded token $\mathbf{x}_i$, yielding
\begin{equation}
\tilde{\mathbf{x}}_i \;=\; \mathrm{Embed}\bigl(\mathbf{x}_i\bigr) \;+\; \mathbf{p}_i,
\end{equation}
where $\mathrm{Embed}$ is the token embedding function. This positional 
encoding is crucial to provide the necessary positional information for the 
self-attention mechanism.

A simple linear decoder projects the final encoded representations back to 
real and imaginary components, reconstructing both the originally masked and 
unmasked portions. 
This masked-reconstruction approach is analogous to masked autoencoders or 
BERT~\cite{BERT}~like masking methods, thereby encouraging the model to 
infer missing entries from the surrounding context.

In our experiments, we tested three configurations corresponding to 
$n=4$ (for $\C^2\otimes \C^2$), $n=6$ (for $\C^2\otimes \C^3$),
and $n=9$ (for $\C^3\otimes \C^3$). All other Transformer hyperparameters 
(embedding dimension, number of attention heads, number of layers, and dropout) 
were selected during the initial research phase by trail-and-error method, and 
remained the same across these experiments.

We used separate datasets for each experiment for pretraining and 
classification, as presented in Table~\ref{tab:training_data}. In every 
experiment, the dataset was divided into training, validation, and test subsets. 
The validation evaluation was performed every epoch, and the test evaluation 
was performed after the training was completed.

\begin{table}
	\setlength{\tabcolsep}{4pt}
	\centering
	\begin{tabular}{llllrrr}
		\toprule
		& \multicolumn{3}{c}{Pretraining} & \multicolumn{3}{c}{Classification} \\
		Group name & $\C^2\otimes \C^2$ & $\C^2\otimes \C^3$ & $\C^3\otimes \C^3$ & $\C^2\otimes \C^2$ & $\C^2\otimes \C^3$ & $\C^3\otimes \C^3$ \\
		\midrule
		sep & 4,000,000 & 8,000,000 & 6,000,000 & 1,000,000 & 1,000,000 & 1,000,000 \\
		general-ent & 2,000,000 & 8,000,000 & 2,000,000 & 300,000 & 1,000,000 & 500,000 \\
		werner-ent & 2,000,000 &  & 2,000,000 & 300,000 &  & 500,000 \\
		max-ent & 2,000,000 &  & 2,000,000 & 300,000 &  & 500,000 \\
		horodecki-bound &  &  & 2,000,000 &  &  & 500,000 \\
		horodecki-ent &  &  & 2,000,000 &  &  & 500,000 \\
		\bottomrule
	\end{tabular}
	
	\caption{Training data sizes for Pretraining and Classifier tasks. 
		Numbers indicate samples used for each model configuration and data 
		type. Group names correspond to the type of data (pure separable,
		general entangled etc.) described in Section~\ref{sec:dataset}. 
		}
	\label{tab:training_data}
\end{table}

\subsection{Pretraining}
In this stage, we train the \emph{MaskedTransformer} from scratch as an 
autoencoder. Specifically, we:
\begin{itemize}
	\item Use the datasets of 10 mln (for $\C^2\otimes \C^2$) and 16 mln ($\C^2\otimes \C^3$ and 
		$\C^3\otimes \C^3$), described in detail in Table \ref{tab:training_data} in the
		\emph{Pretraining} columns.
	\item Split the dataset it into training (90\%), validation 
	(5\%), and test (5\%) partitions.
	\item Randomly mask a given fraction (15\%) of tokens, replacing them with 
	a learned mask token.
	\item Pass the masked tokens through the Transformer encoder and 
	reconstruct them via a linear decoder head.
	\item Optimize the full reconstruction loss using mean squared 
	error (MSE).
\end{itemize}
The training uses a well-known cosine-annealing learning rate schedule and standard PyTorch 
Lightning callbacks for logging and checkpointing. Upon completion, the final 
checkpoints were saved for subsequent classification.

To evaluate the pretraining performace we introduce a metric called Hermitian 
distance, that measures the deviation of a matrix from being perfectly 
Hermitian by computing the average Frobenius norm of the difference between 
the matrix and its conjugate transpose: 

\begin{equation}
h = \frac{1}{b} \sum_{k=1}^{b} \sqrt{||\vect{A}_k-\vect{A}_k^\dagger||_F},
\end{equation}
where $b$ is the number of matrices in the batch, and $\mathbf{A}_k$ denotes 
the $k$-th complex matrix. The notation $\mathbf{A}_k^\dagger$ refers to its 
conjugate transpose, and $\|\cdot\|_F$ indicates the Frobenius norm. For a 
Hermitian matrix, $\mathbf{A}_k = \mathbf{A}_k^\dagger$, which makes the norm 
vanish. When the matrix is split into real and imaginary parts, 
$\mathbf{A}_k = \mathbf{R}_k + i\,\mathbf{I}_k$, Hermiticity requires 
$\mathbf{R}_k$ to be symmetric $\bigl(\mathbf{R}_k = \mathbf{R}_k^T\bigr)$ 
and $\mathbf{I}_k$ to be anti-symmetric 
$\bigl(\mathbf{I}_k = -\mathbf{I}_k^T\bigr)$. Consequently, the quantity 
$\mathbf{A}_k - \mathbf{A}_k^\dagger$ captures deviations from these 
symmetries, and its Frobenius norm measures how far the matrix is from being 
perfectly Hermitian. Averaging over all matrices in the batch yields the 
final distance $h$.

We used this metric to evaluate our pretraining process by assessing how well 
the pretrained model preserves the Hermitian structure of the data  when 15\% 
of the matrix is reconstructed by a network. A lower Hermitian distance 
indicates that the intrinsic mathematical properties are maintained, 
serving as a meaningful indicator of the quality of pretraining.

\subsection{Classifier Training}
After pretraining, we fine-tune or adapt the learned Transformer weights for 
a downstream binary classification task. The core steps are:
\begin{itemize}
	\item Load the pre-trained Transformer weights into a new model that 
		augments the Transformer encoder with a feed-forward classification 
		head (two-class output).
	\item Use the datasets of 1.9 mln (for $\C^2\otimes \C^2$), 2 mln 
		(for $\C^2\otimes \C^3$), and 3.5 mln (for $\C^3\otimes \C^3$), 
		described in detail in Table \ref{tab:training_data} in the 
		\emph{Classification} subsection. This data is separate from the 
		pretraining data.
	\item Split the dataset into training (90\%), validation (5\%), and test 
		(5\%) partitions.
	\item Train the network using cross-entropy loss and the same PyTorch Lightning setup, 
	with logging, checkpointing, and cosine-annealing learning rate schedule.
\end{itemize}
This two-stage approach leverages the pre-trained Transformer’s learned 
representation of the quantum matrices, enhancing the performance of the 
downstream classification task.

\section{Results and discussion}\label{sec:rd}
To determine the final results of the pretraining and classification, we used the
separate data subset that was not used during the training phase neither for
training nor for validation and testing. This subset comprises of 100,000 
samples for each of the classes. The final evaluation was carried out for both 
pretraining and classification after the entire training process was completed.

\paragraph{Pretraining} 
During the pretraining phase, both the loss function and the Hermitian distance 
improved significantly. Table~\ref{tab:hermitian_distances} shows the averaged 
results of the Hermitian distance metric for the pretraining phase. The results
are consistent and show that, overall, the pretraining process ended up 
generating models that are able to reconstruct the Hermitian structure of the
data. Surprisingly, the Hermitian distances achieved very low values during 
the early pretraining. We did not observe any significant improvement in this 
metric after the first few epochs, whereas the loss function continued to 
decrease. This suggests that the model learned the Hermitian structure of 
the data very quickly, and the loss function was optimized to a greater extent.

\begin{table}
\setlength{\tabcolsep}{6pt}
\centering
\begin{tabular}{lcccccc}
\toprule
 & \multicolumn{3}{c}{Untrained} & \multicolumn{3}{c}{Pretrained} \\
Group name & $\C^2\otimes \C^2$ & $\C^2\otimes \C^3$ & $\C^3\otimes \C^3$ & $\C^2\otimes \C^2$ & $\C^2\otimes \C^3$ & $\C^3\otimes \C^3$ \\
\midrule
sep & 6.686 & 3.718 & 8.101 & 0.265 & 0.364 & 0.419 \\
general-ent & 6.704 & 3.716 & 8.098 & 0.189 & 0.187 & 0.167 \\
werner-ent & 6.645 &  & 8.131 & 0.183 &  & 0.222 \\
max-ent & 6.693 &  & 8.102 & 0.296 &  & 0.449 \\
horodecki-bound &  &  & 8.104 &  &  & 0.120 \\
horodecki-ent &  &  & 8.102 &  &  & 0.122 \\
\bottomrule
\end{tabular}

\caption{Averaged Hermitian distances for untrained, and fully pretrained models.}
\label{tab:hermitian_distances}
\end{table}

\paragraph{Classification}
The classification results are presented in Table 
\ref{tab:classification_results}. It is important to emphasize that, while results are presented for each type of state, the classification was deliberately kept binary--entangled vs. separable--as this is the relevant distinction for practical applications.
The results show that the model was able to classify
the states with very high accuracy. The results are consistent across all
dimensions and classes. The only errors appear for the pure separable states 
$\C^2\otimes \C^2$ and $\C^2\otimes \C^3$ groups, where a few of states were 
misclassified as entangled. This confirms the validity of our approach, as the 
model effectively captures the structural properties of quantum states and 
generalizes well across different state classes and dimensions, with only 
minimal misclassification in specific cases.

\begin{table}
\setlength{\tabcolsep}{12pt}
\centering
\begin{tabular}{lllr}
	\toprule
Group name	& $\C^2\otimes \C^2$ & $\C^2\otimes \C^3$ & $\C^3\otimes \C^3$ \\
	\midrule
	sep & 99.995\% & 99.998\% & 100\% \\
	general-ent & 100\% & 100\% & 100\% \\
	werner-ent & 100\% &  & 100\% \\
	max-ent & 100\% &  & 100\% \\
	horodecki-bound &  &  & 100\% \\
	horodecki-ent &  &  & 100\% \\
	\bottomrule
\end{tabular}
	\caption{Accuracy of binary classification (entangled vs separated) for each data group.}
	\label{tab:classification_results}
\end{table}

To further verify our results, we tested whether the deep fine-tuning during
classification training provides an additional learning benefit beyond the 
pretraining phase.
Specifically, we took a pretrained model and froze all its layers except for
the final classification layer, which was then fine-tuned on labeled data.
This approach aimed to determine whether the pretrained representations alone
were sufficient for entanglement classification or if further adaptation was
necessary. The fine-tuned model performed nearly perfectly on 
$\C^2\otimes \C^2$ and $\C^2\otimes \C^3$ states but struggled with 
$\C^3\otimes \C^3$ states, achieving around 85\% accuracy. 
A detailed breakdown revealed that while entangled states were consistently 
classified correctly, the model frequently misclassified separable states. 
This suggests that while the pretrained model captures general entanglement 
patterns, adapting deeper layers during training may be crucial for 
distinguishing subtle features in higher-dimensional separable states.

\paragraph{Discussion} Our work is closely related to~\cite{Goes2021Automated}, 
which also explores automated entanglement classification. However, we achieve 
significantly better results, with near-perfect accuracy compared to their 
reported range of 62–88\%. While their approach struggled with deep learning, 
we successfully adapted transformers into a highly effective classification 
method. An important distinction is that the presence of bound entangled 
states in our dataset does not degrade performance. Additionally, our dataset 
is substantially larger, containing millions of states, whereas theirs 
consisted of only 3,254. Their dataset generation method, in principle, allows 
for bound entangled states in any dimension, while our approach focuses on a 
specific family of states for the $\C^3\otimes \C^3$ case.
However, their generation method 
is much more computationally expensive. Overall, our results are consistent 
with theirs, but we extend the approach significantly, demonstrating the 
feasibility of deep learning for entanglement classification at a much larger 
scale.

Our work serves as an example of the successful integration of machine learning techniques with quantum information science. Similar approaches--whether in developing quantum information models~\cite{Cholewa2017TOM} or solving quantum computing problems ~\cite{Smierzchalski2022ECRL}--represent a promising direction for the field. We expect that this fusion of computational methods and quantum theory will gain increasing prominence, much like the impact of machine learning in computational chemistry.

Our approach addresses a different but related problem compared to~\cite{greenwood2023machine}. Their work focuses on generating entanglement witnesses for specific types of quantum states and specially structured witnesses. While their method is well-executed, it is inherently limited, as reflected in their choice of states. In contrast, our approach is applied to a significantly larger dataset, allowing for a broader and more flexible classification of entanglement. However, their method extends to multipartite entanglement, an area we have not yet explored.

\paragraph{Conclusions}
We have demonstrated that transformer-based neural networks can effectively classify bipartite quantum states as entangled or separable by learning directly from quantum state matrices. By leveraging a masked autoencoding pretraining strategy, our model captures the structural properties of density matrices, achieving near-perfect classification accuracy across various state types and dimensions. These results highlight the potential of modern deep learning architectures for quantum information processing, paving the way for scalable, data-driven approaches to entanglement detection and beyond.

\begin{credits}
\subsubsection{\ackname}
This project was supported by the National Science Center (NCN), Poland, under Projects: Sonata Bis 10, No. 2020/38/E/ST3/00269 (L.P.)

\subsubsection{\discintname}
The authors have no competing interests to declare that are relevant to the
content of this article.
\end{credits}
\bibliographystyle{splncs04}
\bibliography{quantumllm}
\end{document}